\documentclass[pra,twocolumn,showpacs,preprintnumbers,amsmath,amssymb,superscriptaddress]{revtex4}

\usepackage{color}
\usepackage{graphicx}
\usepackage{dcolumn}
\usepackage{bm}
\usepackage{hyperref}
\usepackage{mathrsfs}
\usepackage{soul}

\begin{document}


\title{Split spin-squeezed Bose-Einstein Condensates}

\author{Yumang Jing}
\affiliation{New York University Shanghai, 1555 Century Ave, Pudong, Shanghai 200122, China} 
\affiliation{School of Electronic and Electrical Engineering, University of Leeds, Leeds LS2 9JT, U.K.}

\author{Matteo Fadel}
\affiliation{Department of Physics, University of Basel, Klingelbergstrasse 82, 4056 Basel, Switzerland}

\author{Valentin Ivannikov}
\affiliation{New York University Shanghai, 1555 Century Ave, Pudong, Shanghai 200122, China} 

\author{Tim Byrnes}
\email{tim.byrnes@nyu.edu}
\affiliation{State Key Laboratory of Precision Spectroscopy, School of Physical and Material Sciences,
East China Normal University, Shanghai 200062, China}
\affiliation{New York University Shanghai, 1555 Century Ave, Pudong, Shanghai 200122, China}  
\affiliation{NYU-ECNU Institute of Physics at NYU Shanghai, 3663 Zhongshan Road North, Shanghai 200062, China}
\affiliation{National Institute of Informatics, 2-1-2 Hitotsubashi, Chiyoda-ku, Tokyo 101-8430, Japan}
\affiliation{Department of Physics, New York University, New York, NY 10003, USA}

\date{\today}

\begin{abstract}
We investigate and model the behavior of split spin-squeezed Bose-Einstein condensates (BECs) system. In such a system, a spin-polarized BEC is first squeezed using a $ (S^z)^2 $ interaction, then are split into two separate clouds.  After the split, we consider that the particle number in each cloud collapses to a fixed number.  We show that this procedure is equivalent to applying an interaction corresponding to squeezing each cloud individually plus an entangling operation.  
We analyze the system's entanglement properties and show that it can be detected using correlation-based entanglement criteria. The nature of the states are illustrated by Wigner functions and have the form of a correlated squeezed state.  The conditional Wigner function shows high degrees of non-classicality for dimensionless squeezing times beyond $ 1/\sqrt{N} $, where $N$ is the number of particles per BEC.  
\end{abstract}

            
\maketitle

\section{Introduction}

Entanglement has been traditionally considered to be a fragile quantum phenomenon which only happens in the microscopic world. As the state of the art improves, it has become increasingly realizable in the macroscopic world, and thereby accessible to future quantum technologies.  One of the first demonstrations of entanglement between remote macroscopic objects was achieved between atomic ensembles containing $10^{12}$ atoms, eventually realizing quantum teleportation \cite{julsgaard2001,krauter2011,krauter2013}.  The weak nonlinearity and quantum correlations between fields of light have also been investigated to generate macroscopic entanglement\cite{jeong2005,paternostro2007}. Such entanglement has been considered for being used in various quantum technologies such as metrology and quantum computing  \cite{pezze2016,byrnes12,byrnes15}. In terms of long distance entanglement, the current records for entanglement are using space-based distribution methods, between photons \cite{yin2017,ren2017ground}.   Entanglement in the mesoscopic regime has been experimentally realized by superconductors, which is now considered to be one of the leading candidates for quantum computing \cite{nelson1988,lesovik2001,berkley2003,dicarlo2010}. Micromechanical resonators also exhibit entanglment in the mesoscopic regime, where micrometer-size structures display quantum mechanical behavior \cite{armour2002,genes2008,palomaki2013entangling,groblacher2009observation}.

For Bose-Einstein condensates (BECs), most studies until recently have been focused on creating entanglement within a single atomic cloud.   Spin squeezing, as a means to reduce quantum noise and improve measurement precision beyond the standard quantum limit \cite{wineland1994,appel2009}, can be also applied for the generation of entanglement in BECs \cite{sorensen2001many,riedel2010,gross2012,toth2009}. The creation of many-particle entanglement in one BEC ensemble has been investigated thoroughly \cite{gross2010,gross2012,sorensen2001many,sorensen2001,strobel2014}, which is localized in a single spatial location.  However, it is now considered that entanglement can be furthermore subdivided into various categories, including steerable and non-local entanglement \cite{adesso2016measures}. These categories are in order of increasing quantum correlation, such that non-local entanglement implies both steerability and entanglement, but it is possible to have entangled states that are neither steerable or non-local.  Bell correlations within a single BEC have been observed, showing that the strength of the quantum correlations are of the strongest variety \cite{schmied2016bell}.
Recently, several experiments showing entanglement between spatially separated parts of a single BEC were reported \cite{fadel2017,kunkel2017,lange2017}. To date, there has not been an experimental realization of entanglement between two completely spatially distinct BECs.  Several theoretical studies have analyzed the entanglement created by a $ S^z S^z $ type interaction, pointing to a complex entanglement structure with fractal characteristics \cite{byrnes13,kurkjian2013spin}. 
Numerous proposal have been made for generating such entanglement, based on cold atomic collisions \cite{PhysRevA.74.022312}, atom-light interactions \cite{pyrkov2013,olov,ilo2014theory,rosseau2014,hussain2014}, and Rydberg excitations \cite{idlas2016}. Several proposals for quantum information applications have been made based on such entanglement \cite{byrnes14,ilo2014theory,pyrkov2014quantum}.

\begin{figure}[t]
\includegraphics[width=\columnwidth]{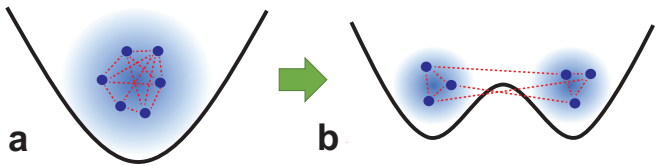}
\caption{\label{conceptual} Schematic procedure for realizing split spin-squeezed Bose-Einstein condensates (BECs). (a) An initial spin coherent state in a single trap is spin-squeezed, producing correlations (dashed lines) between the bosons. (b) The bosons are then spatially separated into two ensembles via a beam splitter interaction. The atoms in the cloud remain correlated, both within and between the two ensembles. }
\end{figure}

In this paper, we analyze another approach for creating entanglement between spatially distinct BECs.  The situation that we consider is shown in Fig. \ref{conceptual}.  First a maximally $S^x$-polarized spin coherent state is squeezed by one-axis twisting, and then it is spatially split into two ensembles. The splitting procedure does not alter the internal state of the atoms, which stay entangled even if they become separated by physical distance. In this work we study the quantum correlations between the two BECs obtained from this protocol, and discuss methods for detecting entanglement using correlations of observables. Recently, Oudot, Sangouard, and co-workers examined entanglement witnesses for such entangled BECs in the presence of local white noise \cite{oudot2017}.  Our physical model differs from this work as we consider the spatial separations to be large enough such that the quantum number fluctuations between split BECs collapses to a fixed number.  This is reasonable from the perspective that quantum tunneling should be highly suppressed for large splitting separations.  
We also analyze the non-classicality and non-Gaussianity of the quantum states that are generated in this system by reconstructing the Wigner functions for the states at various squeezing times.  This helps to visualize the nature of the states to a better degree.  Furthermore, we compare several correlations based methods for detecting entanglement, and find that the Giovannetti {\it et al.} \cite{Giovannetti} and covariance matrix \cite{vinay2017} methods provide a powerful and efficient method for achieving this.  We also show that our system allows for correlations stronger than entanglement as detected by a Einstein-Podolsky-Rosen (EPR) steering criterion.

This paper is structured as follows: In Sec. \ref{ii}, we give a detailed derivation how the process of squeezing and splitting is equivalent to an entangling interaction between two clouds. In Sec. \ref{iii}, we discuss the basic properties of the split spin-squeezed BEC, including the degree of entanglement and its non-Gaussian characteristics represented by the Wigner function. In Sec. \ref{iv} we consider how to detect the entanglement in practice by using several different approaches: the widely used Duan-Giedke-Cirac-Zoller criterion \cite{duan} which is based on the variance of a pair of EPR-type operators, the covariance matrix formalism \cite{simon,guhne,vinay2017}, and the crtierion of Giovannetti {\it et al.} \cite{Giovannetti}. We also show that EPR steering can be observed in the same system in Sec. \ref{sec:steering}. The conclusions are given in Sec. \ref{v}.

\section{Split spin-squeezed Bose-Einstein condensates} \label{ii}


\subsection{The Model}

We now construct a model for the splitting of a spin-squeezed BEC illustrated in Fig. \ref{conceptual}. 
Our model has general similarities to that considered in Ref. \cite{oudot2017}, but with differing treatments of the splitting procedure.  The basic assumptions are that each trap in Fig. \ref{conceptual} can be approximated by two bosonic modes.  Thus before the BEC is split, there are two bosonic modes, consisting of the spatial ground state for the two populated hyperfine states.  After the BEC is split, there are four modes, corresponding to two from each trap.  Further details of this treatment can be found in Refs. \cite{gross2010,byrnes15,byrnes12}.  

First let us introduce some notation.  We denote a general spin coherent state as
\begin{align}
|\alpha ,\beta \rangle\rangle =\frac{1}{\sqrt{N!}}(\alpha a^{\dagger}+\beta b^{\dagger})^N |0\rangle,
\end{align}
where $a^{\dagger}$, $b^{\dagger}$ are bosonic creation operators of the two hyperfine spin states respectively, and $\alpha$, $\beta$ are arbitrary complex coefficients satisfying $|\alpha|^2 + |\beta|^2 = 1$. The Schwinger boson (total spin) operators are defined as 
\begin{align}
S^x &=a^{\dagger}b+b^{\dagger}a, \nonumber\\
S^y &=-ia^{\dagger}b+ib^{\dagger}a, \nonumber\\
S^z &=a^{\dagger}a-b^{\dagger}b,
\end{align}
which obey the commutation relations $[S^l, S^m]=2i\epsilon_{lmn}S^n$, where $\epsilon_{lmn}$ is the Levi-Civita symbol. 

Initially the BEC is prepared in a spin coherent state, where all $N$ atoms occupy the same spin state $ a $.  A $ \pi/2 $ pulse is then applied on the Bloch sphere, such that the state
\begin{align}
| \frac{1}{\sqrt{2}}, \frac{1}{\sqrt{2}} \rangle\rangle =  e^{-i S^y \pi/4 } | 1,0 \rangle\rangle \;,
\end{align}
which is maximally $S^x$-polarized, is prepared.  This is then squeezed using the one-axis twisting evolution, corresponding to
\begin{align}
e^{i(S^z)^2t}| \frac{1}{\sqrt{2}}, \frac{1}{\sqrt{2}} \rangle\rangle = \frac{1}{\sqrt{2^N}} \sum_{k=0}^N \sqrt{N \choose k} e^{i(2k-N)^2 t } | k \rangle 
\end{align}
where
\begin{align}
| k \rangle  =  \frac{(a^{\dagger}) ^{k} (b^{\dagger})^{N-k}}{\sqrt{k ! (N-k) !}} |0\rangle
\end{align}
are the Fock states.  Here $ t $ is a dimensionless time of application of the squeezing operation. The atoms are then spatially separated into two parts. This is described by a beam splitter operation
\begin{align}
a=\frac{1}{\sqrt{2}}(a_L+a_R) \nonumber\\
b=\frac{1}{\sqrt{2}}(b_L+b_R) \label{bs}
\end{align}
where $ a_{L,R} $ and $ b_{L,R} $ are the left and right well modes for the two hyperfine states respectively. We then obtain the state
\begin{align}
|\Psi (t) \rangle = \frac{1}{2^N} &  \sum_{N_L=0}^{N} \sum_{k_L=0}^{N_L} \sum_{k_R=0}^{N - N_L} \sqrt{ \left( \begin{array}{c} N \\ N_L \end{array} \right) \left( \begin{array}{c} N_L \\ k_L \end{array} \right) \left( \begin{array}{c} N - N_L \\ k_L \end{array} \right)} \nonumber\\
                            & \times e^{i(2 k_R + 2 k_L -N)^2 t} |k_L\rangle_{N_L} |k_R\rangle_{N - N_L} , \label{final}
\end{align}
where $N_L$ is the number of atoms in left ensemble.  Here, the eigenstates of $S^z_i$ are
\begin{align}
|k_i\rangle_{N_i} = \frac{(a_i^{\dagger}) ^{k_i} (b_i^{\dagger})^{N_i-k_i}}{\sqrt{k_i ! (N_i-k_i) !}} |0\rangle 
\label{kstatesdef}
\end{align}
where the subscripts take the values $i \in \{L,R \}$ and $N_i $ labels the number of atoms in the BEC.

The wavefunction (\ref{final}) has a sum over $ N_L $, the number of particles in the left well.  This means that the particle number per well is not fixed, and exist in a superposition across both wells.  However, for completely spatially separated BECs, it may be challenging to realize atoms delocalized over large distances.  For example,  there might be some inadvertent leakage of the particle number information which will collapse the superposition to a particular $ N_L $ and $ N_R $ on the left and right wells respectively. At the very least, there will always be a collapse during the measurement process, which will simultaneously measure particle number as well as spin.  We thus assume that there is additionally some decoherence which can be taken into account by a projection (number fixing measurement) onto a left-well particle number $ N_L $ performed by the operator
\begin{align}
P_{N_L}= \left( \sum_{k_L=0}^{N_L} | k_L \rangle_{N_L} \langle k_L |_{N_L}  \right)
\left( \sum_{k_R=0}^{N_R} | k_R \rangle_{N_R} \langle k_R |_{N_R}  \right) ,
\end{align}
where the number of atoms in the right well is
\begin{align}
N_R = N - N_L .
\end{align}
Here the terms in the brackets give the identity operation in a fixed subspace of particle number for the left and right wells. After being projected, the normalized state can be expressed as 
\begin{align}
|\Psi^{N_L} (t) \rangle &= \frac{P_{N_L} | \Psi (t) \rangle }{\sqrt{p(N_L)}} \nonumber \\
& =  \frac{1}{\sqrt{2^N}} \sum_{k_L=0}^{N_L} \sum_{k_R=0}^{N_R} \sqrt{ \left( \begin{array}{c} N_L \\ k_L \end{array} \right) \left( \begin{array}{c} N_R \\ k_R \end{array} \right)} \nonumber\\
                            & \times e^{i(2 k_R + 2 k_L -N)^2 t} |k_L\rangle_{N_L} |k_R\rangle_{N_R} 
														\label{measuredstate}
\end{align}
where the probability of obtaining the state with $ N_L $ atoms in the left well is given by
\begin{align}
p(N_L) & = \langle \Psi (t) | P_{N_L}^2 | \Psi (t) \rangle \nonumber \\
& = \frac{1}{2^N} \left( \begin{array}{c} N \\N_L \end{array} \right) .
\label{leftwellprob}
\end{align}
It is illuminating to write (\ref{measuredstate}) in the form
\begin{align}  
|\Psi^{N_L} (t) \rangle  &=e^{i(S_L^z+S_R^z)^2t} | \frac{1}{\sqrt{2}}, \frac{1}{\sqrt{2}}\rangle\rangle_{N_L} | \frac{1}{\sqrt{2}}, \frac{1}{\sqrt{2}} \rangle\rangle_{N_R}  \label{sub} \\
& = \frac{e^{i[ (S_L^z)^2 +(S_R^z)^2] t} }{\sqrt{2^{N_R}}} \sum_{k_R=0}^{N_R} \sqrt{N_R \choose k_R}  \nonumber \\
& \times | \frac{e^{2i(2k_R-N_R)t}}{\sqrt{2}}, \frac{e^{-2i(2k_R-N_R)t}}{\sqrt{2}} \rangle \rangle_{N_L} | k_R \rangle_{N_R} .
\label{illuminating}
\end{align}
which shows that the final state can be viewed to have a fixed particle number in each well
acted on by the Hamiltonian
\begin{align}
H_{\text{eff}} =(S_L^z+S_R^z)^2 = (S_L^z)^2 + (S_R^z)^2 + 2 S_L^z S_R^z  .
\label{effectiveham}
\end{align}
This makes it clear that the split squeezing operation is equivalent to a combination of squeezing each of the split BECs individually, as well as applying the $ S_L^z S_R^z $ operation, which creates entanglement between the two BECs \cite{byrnes13,kurkjian2013spin}.

The state (\ref{measuredstate}) corresponds to the conditional state for a particular collapse of the atom number, to $ N_L $ atoms in the left well. Experimentally, this would correspond to post-selecting the measurement outcome with $ N_L $ atoms in the left well.  After many runs of the experiment, the state that is measured is a probabilistic mixture of various outcomes.  The mixed state that is obtained is 
\begin{align}
\rho & = \sum_{N_L=0}^N p(N_L)  |\Psi^{N_L} (t)  \rangle  \langle  \Psi^{N_L} (t) |  \label{fixfinal}
\end{align}
We will examine both the conditional state (\ref{measuredstate}) and the average ensemble state (\ref{fixfinal}) to examine their properties in Sec. \ref{iii}.

\subsection{Equivalence of split-squeezing with squeeze-splitting}

\label{sec:equiv}

In (\ref{illuminating}) we found that the procedure of splitting a squeezed BEC was equivalent to splitting a condensate and then applying a squeezing on the total spin of the system (\ref{effectiveham}).  On first glance this is puzzling because the spin operators $ S_{L,R}^z $ do not commute with the splitting operation (\ref{bs}). To show that in fact the order of these operations do not matter, let us first define the splitting operators
\begin{align}
\Sigma_a^x & = a_L^\dagger a_R +   a_R^\dagger a_L \nonumber \\
\Sigma_a^y & = -i a_L^\dagger a_R +  i a_R^\dagger a_L \nonumber \\
\Sigma_b^x & =  b_L^\dagger b_R +   b_R^\dagger b_L \nonumber \\
\Sigma_b^y & = -i b_L^\dagger b_R +  i b_R^\dagger b_L .
\end{align}
The operation of (\ref{bs}) then amounts to the transformation 
\begin{align}
e^{i \Sigma_a^y \pi/4} a_R e^{-i \Sigma_a^y \pi/4} & = \frac{1}{\sqrt{2}} ( a_L + a_R) \nonumber \\
e^{i \Sigma_b^y \pi/4} b_R e^{-i \Sigma_b^y \pi/4} & = \frac{1}{\sqrt{2}} ( b_L + b_R) .
\end{align}
Thus in this formulation we identify the original modes as left-well operators $ a \equiv a_L $, $ b \equiv b_L $, and these are split by the beam-splitter operation. The state (\ref{measuredstate}) can be thus written equivalently as
\begin{align}
| \Psi^{N_L} (t) \rangle = \frac{P_{N_L} e^{i \Sigma_a^y \pi/4} e^{i \Sigma_b^y \pi/4} e^{i(S^z_R)^2 t}| \frac{1}{\sqrt{2}}, \frac{1}{\sqrt{2}} \rangle\rangle_R}{\sqrt{p(N_L)}}  ,
\label{rewrittenpsi}
\end{align}
where the initial state state is written in terms of the left-well operators $ a_L, b_L $.  

The question here is whether the squeezing operation can be put after the splitting operation.  Simply inverting the $ S^z_R $ and $ \Sigma_{a,b}^y $ operations is clearly not allowed since they do not commute:
\begin{align}
[S^z_R, \Sigma^y_a] & = i \Sigma^x_a \nonumber \\
[S^z_R, \Sigma^y_b] & = -i \Sigma^x_b .
\end{align}
The key point to notice is that since
\begin{align}
[S^z_L, \Sigma^y_a] & = -i \Sigma^x_a \nonumber \\
[S^z_L, \Sigma^y_b] & = i \Sigma^x_b ,
\end{align}
the combination of $ S^z_L + S^z_R $ does commute with the splitting operators
\begin{align}
[ S^z_L + S^z_R, \Sigma^y_a] = [ S^z_L + S^z_R, \Sigma^y_b] =0  .
\label{commutingtotalspin}
\end{align}

Using this we can directly show the desired relation.  Since there are no right-well operators in the state (\ref{rewrittenpsi}), we can introduce a additional factor of $ e^{i[ (S^z_L)^2 + 2 S^z_L S^z_R ]t} $ just before the state, giving
\begin{align}
| \Psi^{N_L}\rangle =\frac{ P_{N_L}   e^{i \Sigma_a^y \pi/4} e^{i \Sigma_b^y \pi/4} e^{i(S^z_L+ S^z_R)^2 t}| \frac{1}{\sqrt{2}}, \frac{1}{\sqrt{2}} \rangle\rangle_R}{\sqrt{p(N_L)}} ,
\label{rewrittenpsi2}
\end{align}
Now using (\ref{commutingtotalspin}), we can commute the squeezing operation to the end, giving 
\begin{align}
| \Psi^{N_L} \rangle = \frac{e^{i(S^z_L+ S^z_R)^2 t} P_{N_L}  e^{i \Sigma_a^y \pi/4} e^{i \Sigma_b^y \pi/4} | \frac{1}{\sqrt{2}}, \frac{1}{\sqrt{2}} \rangle\rangle_R}{\sqrt{p(N_L)}} ,
\label{rewrittenpsi3}
\end{align}
where we used the fact that 
\begin{align}
[ P_{N_L}, S^z_L] =[ P_{N_L}, S^z_R]  = 0
\end{align}
to commute the squeezing operation to the end.  This shows the desired relation.

\section{Basic properties of the state} 
\label{iii}

We now examine the basic properties of the conditional state (\ref{measuredstate}) and the averaged state (\ref{fixfinal}). These can be efficiently characterized by  entanglement between the left and right wells, and the Wigner functions representing the states.  These are examined in the following sections.

\subsection{Entanglement}

\begin{figure}
\includegraphics[width=\columnwidth]{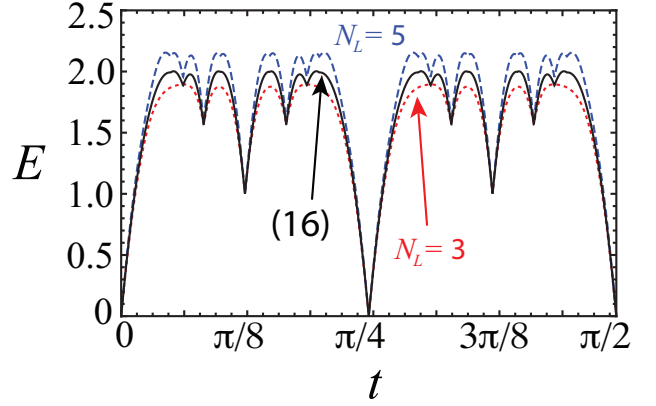}
\caption{Entanglement in split squeezed Bose-Einstein condensates, as quantified by the logarithmic negativity. Solid line corresponds to the state (\ref{fixfinal}), the dashed line and dotted lines represent the conditional state (\ref{sub}) with $N_L=5$ and $N_L=3$ respectively. The total number of atoms is $N=10$.
\label{devil} }
\end{figure}

As a measure of the entanglement between the left and right wells, we use the logarithmic negativity\cite{vidal2002,plenio2005} defined as 
\begin{align}
E (\rho) = \log_2 || \rho^{T_L}||
\end{align}
where $\rho^{T_L}$ is the partial transpose of (\ref{fixfinal}) with respect to left well and $||X|| = \text{Tr} |X| = \text{Tr} \sqrt{X^{\dagger}X}$ is the trace norm of an operator $X$. For two BECs each with two components with particle numbers $ N_L $ and $ N_R $, a maximally entangled state is
\begin{align}
|\Psi_{\text{max}} \rangle = \sum_{k = 0 }^{\min(N_L, N_R)} | k \rangle_{N_L} | k \rangle_{N_R}  .
\label{maxentangled}
\end{align}
This has an logarithmic negativity of
\begin{align}
E (|\Psi_{\text{max}} \rangle  \langle \Psi_{\text{max}}  |)  =  \log_2 [ \min( N_L,N_R) +1 ] .
\label{maxent}
\end{align}
We note that (\ref{maxentangled}) is only one out of $(N+1)^2 $ maximally entangled states.  For example, there are 4 orthogonal Bell states in the case of two qubits which are all maximally entangled.  The remaining states can be generated by local unitary transformations.  

Figure \ref{devil} shows the entanglement as a function of the dimensionless squeezing time for both the conditional state (\ref{measuredstate}) and the averaged state (\ref{fixfinal}).  The first thing that is evident is that all the curves have a similar form to the ``devil's crevasse'' entanglement as seen for a $ S^z_L  S^z_R $ interaction \cite{byrnes13,kurkjian2013spin}.   The entanglement increases monotonically until the characteristic time $t=1/2\sqrt{N}$, with a overall periodicity of $T = \pi/4$ and dips occurring at time that are a fractional multiple of the this time, in agreement to a pure  $ S^z_L  S^z_R $ interaction. 
This is the expected behavior considering that the final interaction can be written in the form (\ref{sub}) which is a combination of this interaction and local squeezing on each well.  Since entanglement is invariant under local transformations, the effect of the local squeezing is not visible in the curves shown in Fig. \ref{devil}.  The dominant effect of the different partitions of the particle number between the wells is to change the amplitude of the entanglement.  This can be understood from the relation (\ref{maxent}) which shows that the amount of entanglement is related to the dimensionality of the minimum of the particle number between the left and right well.  The curves in Fig. \ref{devil} never reach the maximal value of entanglement ($\log_2(N/2+1)\approx 2.6$ here), as the type of entangled states that are generated from a $ S^z_L  S^z_R $ interaction do not given the  maximum value (\ref{maxent}) \cite{byrnes13}.  As expected, 
the amplitude of the entanglement for the case $ N_L = N_R = 5 $ is the largest with respect to all other $N$ partitions, and is symmetric around this midpoint.  The entanglement is well-preserved despite the mixing between all possible partition of $N$ in the state (\ref{fixfinal}).  This agrees with the general arguments of Sec. \ref{sec:equiv}, where the entangling operation commutes with the splitting and localization operations.

\subsection{Wigner functions}

The results of the previous section show that the split squeezed states are entangled, but do not clearly show the nature of the states, including what kind of quantum correlations are present between the two BECs.  To visualize the state it is beneficial to calculate the Wigner functions \cite{wigner,dowling94,schmied2011}, defined for a two mode BEC. The Wigner function gives a representation of the quantum state as a quasiprobability distribution on the Bloch sphere.  
The Wigner function is defined as
\begin{align}
W(\theta, \varphi) = \sum_{k=0}^{2j} \sum_{q=-k}^{k} \rho_{kq} Y_{kq} (\theta, \varphi) \label{wigner}
\end{align}
where $ Y_{kq} (\theta, \varphi)$ are the spherical harmonics.  Here, $\rho_{kq}$ is defined as 
\begin{align}
\rho_{kq}= \sum_{m=-j}^{j} \sum_{m'=-j}^{j} &   (-1)^{j-m} \sqrt{2k+1} \left( \begin{array}{ccc} j & k & j \\ -m & q & m' \end{array} \right) \nonumber \\
& \times \langle jm |\rho | jm' \rangle
\end{align}
where $\left( \begin{array}{ccc} j & k & j \\ -m & q & m' \end{array} \right)$ is the Wigner $3j$ symbol. In the above, we have used notation conventional to the Dicke representation in terms of angular momentum eigenstates. This can be related to our previous notation according to the correspondence
\begin{align}
| j m \rangle = | k =j+m \rangle_{N = 2j} ,
\end{align}
where the left hand side are the Dicke state and the right hand side is defined according to (\ref{kstatesdef}).

The above representation allows us to represent any state of a two component BEC with fixed particle number as a Wigner distribution.  In our case, we have two BECs, each with two components.  This would result in general as a higher dimensional Wigner distribution which is difficult to visualize.  For this reason we consider the marginal and conditional Wigner function respectively \cite{olov}. For the marginal Wigner function, the density matrix for one of the BECs is traced over, and the Wigner function is computed from the remaining density matrix.  For the conditional Wigner function, a projection onto a specific Fock state $|k \rangle$ is made on one of the BECs, and normalized accordingly.  In either case, the resulting Wigner functions only have two variables which can be visualized in a two-dimensional phase space corresponding to the Bloch sphere. A rough intuition for the difference between the two types distribution is that the marginal Wigner function gives an average state of one of the BECs disregarding the state of the other BEC.  The conditional Wigner function gives the correlations that are present given a particular measurement $|k \rangle$ (i.e. post-selection) on the other BEC.  A similar argument was used to construct a visualization for the states in Ref. \cite{byrnes13}.

\subsubsection{Marginal Wigner function}

The reduced density matrix on the left well is
\begin{align}
\rho_L = \text{Tr}_R (  |\Psi^{N_L} (t) \rangle \langle \Psi^{N_L} (t) | ) \label{pt} ,
\end{align}
Inserting this state into (\ref{wigner}), we obtain the marginal Wigner function
\begin{align}
& W  (\theta, \varphi) = \frac{1}{2^N} \sum_{k=0}^{N_L} \sum_{q=-k}^{k} Y_{kq}(\theta, \varphi) \sum_{m_L, m_L^{'}=-N_L/2}^{N_L/2} 
 \sqrt{2k+1} \nonumber\\
& \times \sqrt{\left( \begin{array}{c} N_L \\ N_L/2+m_L^{'} \end{array} \right)  \left( \begin{array}{c} N_L \\ N_L/2+m_L \end{array} \right) } \left( \begin{array}{ccc} N_L/2 & k & N_L/2 \\ -m_L & q & m_L^{'} \end{array} \right)\nonumber\\
                       &\times   e^{i4(m_L-m_L^{'})(m_L+m_L^{'}-N_R)t}  (1+e^{i8(m_L-m_L^{'})t})^{N_R}   (-1)^{j_L-m_L} 
\end{align}
where the total spin on the left well was taken to be a fixed value  $j_L=\frac{N_L}{2}$.

\begin{figure}[t]
\includegraphics[width=\columnwidth]{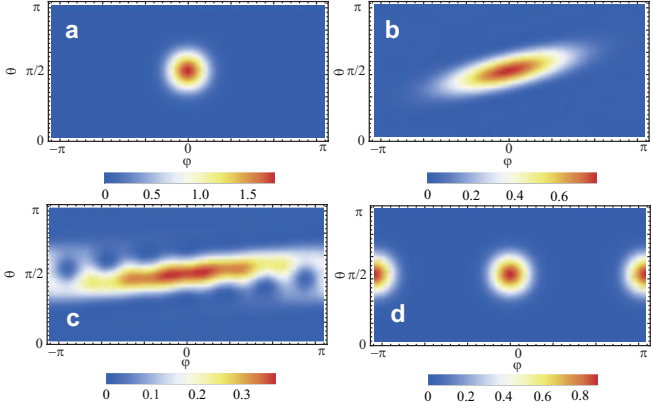}
\caption{\label{marginal}  Marginal Wigner function for the split spin-squeezed state at various interaction times. (a) $t=0$, (b) $t=1/N$, (c) $t=1/2\sqrt{N}$, (d) $t=\pi/8$. In all plots, the total number of atoms is $N=20$, and the numbers of atoms in each ensemble are $N_L=N_R=10$.}
\end{figure}

In Fig. \ref{marginal}, we see the behavior of marginal Wigner function for different times. Initially, at $t=0$, the state is a spin coherent state located at $\theta=\pi/2, \varphi=0$. At short evolution times $0< t \lesssim 1/2\sqrt{N}$, the distribution elongates in a diagonal direction reminiscent of a one-axis squeezing interaction $ (S^z)^2 $.  The width along the squeezing direction is however broader than what would be obtained from a genuine one-axis twisting dynamics, due to the averaging obtained by tracing over the BEC in the right well. After the characteristic time $t\sim 1/2\sqrt{N}$, the state transforms into a non-Gaussian state, where the elongation starts to wrap around the whole Bloch sphere.  At $t=\pi/8$, the state becomes a Bell state consisting of a Schrodinger cat state which is located at  $\varphi=0$ and $\varphi=\pi$ respectively. At this point the state can be written 
\begin{align}
& | \Psi^{N_L}(t=\pi/8)\rangle  = \frac{e^{i\pi/4}}{\sqrt{2}} \Big( | \frac{1}{\sqrt{2}}, \frac{1}{\sqrt{2}} \rangle \rangle_L | \frac{1}{\sqrt{2}}, \frac{1}{\sqrt{2}} \rangle \rangle_R \nonumber \\
& - i (-1)^{(N_L+N_R)/2}
| \frac{-1}{\sqrt{2}}, \frac{1}{\sqrt{2}} \rangle \rangle_L | \frac{-1}{\sqrt{2}}, \frac{1}{\sqrt{2}} \rangle \rangle_R \Big),
\end{align}
which is only valid for even $N_L, N_R$.  Tracing out the BEC in the right well thus gives the state
\begin{align}
\rho_L(t=\pi/8)=
 & \frac{1}{2} \Big( | \frac{1}{\sqrt{2}}, \frac{1}{\sqrt{2}} \rangle \rangle_L \langle \langle \frac{1}{\sqrt{2}}, \frac{1}{\sqrt{2}}  |_L \nonumber \\
 & +
 | \frac{-1}{\sqrt{2}}, \frac{1}{\sqrt{2}} \rangle \rangle_L \langle \langle \frac{-1}{\sqrt{2}}, \frac{1}{\sqrt{2}}  |_L \Big) ,
\end{align}
which is a mixture of the states at opposite ends of the Bloch sphere, matching with the distribution in Fig. \ref{marginal}(d).

\subsubsection{Conditional Wigner function}

The conditional Wigner function is obtained by projecting the final state (\ref{measuredstate}) onto different $|k_R\rangle$ states, and is written 
\begin{align}
P_{k_R} &|\Psi^{N_L} (t) \rangle  = \frac{1}{\sqrt{2^{N_R}}} \sqrt{N_R \choose k_R} e^{i [(S_L^z)^2+(S_R^z)^2]  t}  \nonumber \\
& \times  | \frac{e^{2i(2k_R-N_R)t}}{\sqrt{2}}, \frac{e^{-2i(2k_R-N_R)t}}{\sqrt{2}} \rangle \rangle_{N_L} | k_R \rangle_{N_R}
  \label{rhoj}
\end{align} 
where $P_{k_R} = |k_{r}\rangle_{N-N_L} \langle k_{r}|_{N-N_L} $ is the projector onto Fock states of the right well and we used (\ref{illuminating}). Together with (\ref{rhoj}) and (\ref{wigner}), we thus obtain 
\begin{align}
& W_{k_R} (\theta, \varphi) = \frac{1}{\mathscr{N}} \sum_{k=0}^{N_L} \sum_{q=-k}^{k} Y_{kq}(\theta, \varphi) \sum_{m_L, m_L^{'}=-N_L/2}^{N_L/2}    \sqrt{2k+1} \nonumber \\
& \times \sqrt{ \left( \begin{array}{c} N_L \\ N_L/2+m_L \end{array} \right) \left( \begin{array}{c} N_L \\ N_L/2 +m_L^{'} \end{array} \right)} \left( \begin{array}{ccc} N_L/2 & k & N_L/2 \\ -m_L & q & m_L^{'} \end{array} \right) \nonumber \\
& \times \left( \begin{array}{c} N_R \\ k_R \end{array} \right) e^{i4(m_L-m_L^{'})(2k_R-N_R+m_L+m_L^{'})t}  (-1)^{N_L/2-m_L} 
\end{align}
where $\mathscr{N}$ is the normalization factor and again we took $j_L=\frac{N_L}{2}$.

\begin{figure}[t]
\includegraphics[width=\columnwidth]{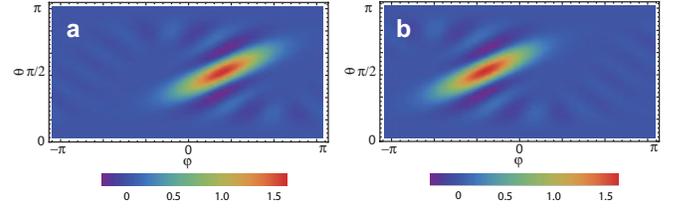}
\caption{\label{joint_kr}  Conditional Wigner function for the split spin-squeezed state after being projected onto various $|k_R\rangle$ states at $t=1/N$ for (a) $k_R=3$ and (b) $k_R=7$. The $ k_R = 5 $ case is plotted in Fig. \ref{joint_time}(a).  
In all plots, the total number of atoms is $N=20$, and the numbers of atoms in each ensemble are $N_L=N_R=10$.}
\end{figure}

\begin{figure}[t]
\includegraphics[width=\columnwidth]{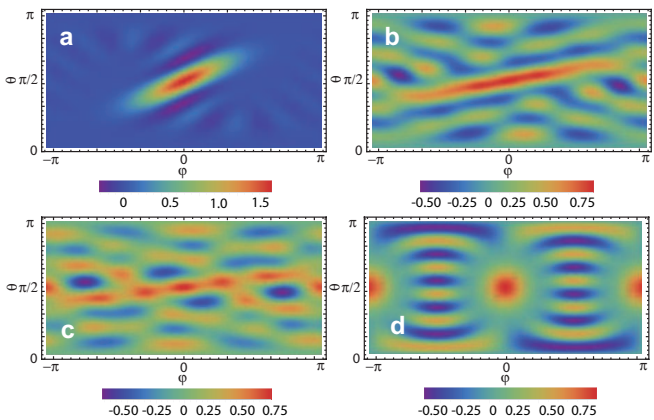}
\caption{\label{joint_time}  Conditional Wigner function for the split spin-squeezed state at various interaction times with $k_R=5$. (a) $t=1/N$, (b) $t=1/\sqrt{2N}$, (c) $t=1/\sqrt{N}$, (d) $t=\pi/8$. In all plots, the total number of atoms is $N=20$, and the numbers in each ensemble are $N_L=N_R=10$.}
\end{figure}

Figure \ref{joint_kr} and Fig. \ref{joint_time}(a) shows the conditional Wigner functions for various projected states $| k_R \rangle $ for the right well. We see that for all the projected values the Wigner functions take the form of one-axis squeezing $ (S^z)^2 $, with the characteristic diagonal distribution.  Unlike the marginal Wigner function, the width of the distributions are narrowed, displaying genuine squeezing.  This can be attributed to the $ (S_L^z)^2 $ term in (\ref{rhoj}) which produces the squeezing effect. For various projected values of $ k_R $, the distributions are offset by various positions around the Bloch sphere, as can be seen by the arguments of the spin coherent state in (\ref{rhoj}). This arises because of the $ S_L^z S_R^z $ interaction which produces an entangled state in the Fock states of the right well and rotated spin coherent states, as shown in (\ref{illuminating}). This is exactly the type of correlations that are expected with this type of interaction and have been discussed in detail in Ref. \cite{byrnes13}.  Thus the integrated density corresponds to the probability of a particular measurement result $ k_R $ is obtained.  The probability distribution is 
\begin{align}
P(k_R)=\text{Tr} (\rho_{N_l} P_{k_R}) = \frac{1}{2^N} \left( \begin{array} {c} N_L \\k_R \end{array} \right) \left( \begin{array} {c} N_R \\k_R \end{array} \right) .
\label{probabilitydistkp}
\end{align}

In Fig. \ref{joint_time}, we show the conditional Wigner distributions for the state (\ref{rhoj}) for various evolution times projected at $k_R=N_R/2$ with time evolution.  The $k_R=N_R/2$ projection is chosen because it is the most probable result according to (\ref{probabilitydistkp}). The behavior of the conditional Wigner function is consistent with that of marginal Wigner function. The initial spin coherent state evolves into a squeezed state at short times followed by a non-Gaussian state for longer time evolution. As the squeezing time $t$ increases, the volume of the negative part of Wigner function is increased, which indicates the emergence of non-classical  behavior \cite{kenfack2004}. This is in contrast with the marginal Wigner function, which is always positive  as was seen in Fig. \ref{marginal}.  According to Hudson's theorem, only pure quantum states with non-negative Wigner functions shows Gaussianity \cite{gross2006,mandilara2009}. Therefore the positivity of Wigner function in this case does not necessarily mean the quantum state is always Gaussian. The conditional Wigner function, corresponding to postselecting certain measurement outcomes, exhibits more non-Gaussianity and higher degree of quantumness.

\section{Detection of entanglement} \label{iv}

As we have already seen in Fig. \ref{devil}, entanglement between the two wells of the split squeezed BEC is always present except for times $ t = 0 $ modulo $ T = \pi/4 $.  This was shown using the logarithmic negativity, which in general requires the full density matrix for the calculation.  Experimentally, this is not convenient as full tomography of the density matrix is required to evaluate the expression.  For example, in Ref. \cite{fadel2017} the total number of atoms was $ N \approx 500 $.  Typically in an experiment, only low order expectation values of the total spins are observable.  Thus the use of correlation-based entanglement criteria \cite{duan, hz2006,Giovannetti,simon,guhne,vinay2017}, where a small number of observables are measured is desirable. In this section, we compare several types to see which gives the most sensitive detection of entanglement.

\subsubsection{DGCZ entanglement criterion}

We first calculate a DGCZ criterion for the split squeezed BEC.  Since the spins are initially polarized in the $ S^x $-direction, we may use a Holstein-Primakoff transformation to treat the other spins as approximate position and momentum operators
\begin{align}
X_i \approx \frac{S^y_i}{\sqrt{2N_i}} \nonumber \\
P_i \approx \frac{S^z_i}{\sqrt{2N_i}}
\end{align}
where $ i \in \{ L, R \} $ and $ S^x_i \approx N_i $ is treated classically. As derived in Ref. \cite{byrnes13}, these variables can be used to create  EPR-like variances which we expect to have suppressed noise fluctuations.  Any separable state then obeys the inequality
\begin{align}
\mathcal{E}_{\text{D}} \equiv \frac{\langle \Delta^2 [ S_L^y - S_R^z ] \rangle + \langle \Delta^2 [ S_R^y - S_L^z] \rangle}{2\langle S_L^x \rangle + 2\langle S_R^x \rangle} \ge 1  \label{duan}
\end{align}
where the variance of an operator  $ A $ is $ \langle \Delta^2 A \rangle = \langle A^2 \rangle -  \langle A \rangle^2 $.  A violation of (\ref{duan}) signals the presence of entanglement. 

The criterion (\ref{duan}) only detects entanglement in a limited time range.  The range can be estimated by evaluating the Heisenberg equations of motion
\begin{align}
S_L^y(t) & \approx S_L^y(0) + 4 N_L t ( S_L^z(0) + S_R^z(0) )  \nonumber \\
S_R^y(t) & \approx S_R^y(0) + 4 N_R t ( S_L^z(0) + S_R^z(0) ) \nonumber \\
S_L^z(t) & = S_L^z(0) \nonumber \\
S_R^z(t) & = S_R^z(0) 
\label{heisen}
\end{align}
Substituting this into (\ref{duan}) for the initial state of completely polarized spins in the $ S^x_i $-direction, we obtain
\begin{align}
\langle \Delta^2 [ S_L^y - S_R^z ] + \langle \Delta^2 [ S_R^y - S_L^z] \rangle \approx 2N ( 4 N^2 t^2 - 2N t + 1)
\label{lhsestimate}
\end{align}
Here, we have approximated that $ N_L = N_R = N/2 $, which is the most likely outcome of (\ref{leftwellprob}).  Similarly, we have for the denominator of (\ref{duan}):
\begin{align}
2\langle S_L^x \rangle + 2\langle S_R^x \rangle = 2N .
\label{rhsestimate}
\end{align}
Thus we expect the entangled states to be detected in the region $ 0 < t < 1/2N $.

Evaluation of the criterion (\ref{duan}) is shown in Fig. \ref{criteriaDvsG}. We see that entanglement can be detected up to times $ t \approx 1/2N $ as expected.  We note that the plot for Fig. \ref{criteriaDvsG} is calculated using the exact expressions for the quantities in (\ref{lhsestimate}) and (\ref{rhsestimate}), rather than the Holstein-Primakoff approximation as above.  We attribute the failure of the criterion (\ref{duan}) beyond times $ t > 1/2N $ to the break down of the  Holstein-Primakoff approximation,  since the total spin is no longer polarized in the $ S^x $ direction after this time, as also can be observed from Fig. \ref{marginal}.

\subsubsection{Covariance matrix entanglement criterion}

Another correlation-based approach that can be used to detect entanglement is the covariance matrix formalism \cite{simon2000peres}. Using Holstein-Primakoff approximated variables, one can construct a $ 4 \times 4 $ covariance matrix in operators $ (X_L, P_L, X_R, P_R) $ which can be used to construct a entanglement criterion.  In order to detect entanglement in a wider time range it is however desirable to not explicitly rely upon the Holstein-Primakoff approximation which we expect to break down for times $ t > 1/N $. 
Recently, the covariance matrix procedure was generalized such that an arbitrary set of operators $ \xi_n $ could be used to construct a covariance matrix \cite{vinay2017}.  In the procedure given in Ref. \cite{vinay2017}, the covariance matrix is constructed for a particular state
\begin{align}
V_{nm} \equiv  \frac{1}{2} \langle \{ \Delta \xi_n, \Delta  \xi_m \} \rangle,
\end{align}
where $ \Delta \xi_n = \xi_n - \langle \xi_n \rangle $.  Then combining this with the commutation matrix
\begin{align}
\Omega_{nm} \equiv -i \langle [ \xi_n, \xi_m ] \rangle,
\end{align}
for any separable state we have
\begin{align}
\text{PT}(V) + \frac{i}{2} \text{PT} (\Omega) \geq 0, \label{covariance}
\end{align}
where $ \text{PT} $ denotes a partial transposition operation. As (\ref{covariance}) is a matrix equation, the meaning of the inequality is in the semi-positive definite nature of the matrix on the left hand side. This means that any negative eigenvalue of $ \text{PT}(V) + \frac{i}{2} \text{PT} (\Omega)  $ signals the violation of the separability condition.   

In our case, the set of observables that we will use are
\begin{align}
\xi = (S_L^x, S_L^y, S_L^z, S_R^x, S_R^y, S_R^z)
\label{xiset}
\end{align}
In Fig. \ref{criteriaDvsG} we plot the quantity
\begin{align}
\mathcal{E}_{\text{CM}} = \frac{\min_i \lambda_i}{N} ,
\label{covariancemat}
\end{align}
where $ \lambda_i $ are the eigenvalues of the left hand side of (\ref{covariance}). We see that this criterion also detects entanglement in the split squeezed BEC state, but for a wider range of times than using the DGCZ criterion.  The region of applicability of the covariance matrix is found to be up to times $ t \approx 1/\sqrt{12 N} $ which we find empirically using several values of $ N $.  In this sense, we find that the covariance matrix  formalism is more powerful than the DGCZ criterion as it can detect entanglement in a wider range.  This is natural since more information is contained in the covariance matrix, since (\ref{xiset}) giving all the  $ 6 \times 6 $ correlators, rather than the two types in (\ref{duan}).

\subsubsection{Giovannetti \textit{et al.} entanglement criterion}

Another entanglement criterion we consider is the one proposed in Ref. \cite{Giovannetti} by Giovannetti \textit{et al.}.  This is given by
\begin{align}
\mathcal{E}_{\text{G}} & \equiv \dfrac{ \sqrt{ \langle \Delta^2 [ g^{z} S_L^{z^\prime} - S_R^{z^\prime} )] \rangle \langle \Delta^2 [ g^{y} S_L^{y^\prime} - S_R^{y^\prime} )] \rangle  } }{ \vert g^{z} g^{y^\prime} \vert \langle S_L^x \rangle + \langle S_R^x \rangle  } 
& \geq 1 \;, \label{giovannetti}
\end{align}
where 
\begin{align}
S_i^{z^\prime} & = S_i^y \sin \theta  + S_i^z \cos \theta  \nonumber \\
S_i^{y^\prime} & = S_i^y \cos \theta  - S_i^z \sin \theta 
\end{align}
are spin operators that are chosen to minimize the variance of $ S_i^{z^\prime} $ and $S_i^{y^\prime} $ for $ i  \in \{L, R \} $.  The squeezing angle is given by 
\begin{align}
\tan 2 \theta = \frac{4 \sin(4t) \cos^{N-2} (4t)}{1- \cos^{N-2} (8 t)} .
\end{align}
The parameters $ g^{y,z} $ are free real parameters that are chosen to minimize $\mathcal{E}_{\text{G}}$. As for the case with (\ref{duan}), the inequality (\ref{giovannetti}) is valid for separable states. A violation of (\ref{giovannetti}) signals the presence of entanglement. The criterion  (\ref{giovannetti}) was successfully adopted in Ref. \cite{fadel2017} to detect entanglement between two regions of an expanded spin-squeezed BEC with $N\approx 600$.

\begin{figure}
\includegraphics[width=\columnwidth]{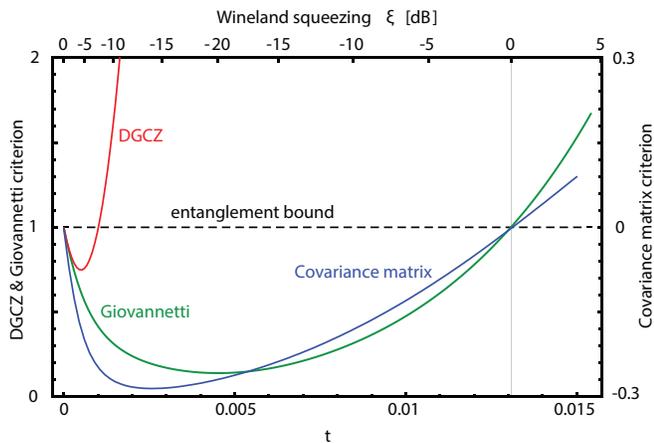}
\caption{Comparison of the DGCZ (\ref{duan}), covariance matrix (\ref{covariancemat}), and Giovannetti (\ref{giovannetti}) entanglement criteria for a split squeezed Bose-Einstein condensate for $N=500$.  The horizontal dashed line is the separability bound, below which entanglement is detected. The top scale shows the Wineland squeezing parameter (\ref{Wineland}), which initially decreases with $ t $, but then increases again. The vertical gray line indicates the point where $\xi=1$. 
\label{criteriaDvsG}
}
\end{figure}

In Fig. \ref{criteriaDvsG} we plot the criteria (\ref{giovannetti}). It is evident that the criterion (\ref{giovannetti}) allows to detect entanglement for a wider range of $t$ than the DGCZ criterion and the same range as the covariance matrix approach.  We have verified that the covariance matrix and Giovannetti criterion give the same range of entanglement detection for any value of $ N $ chosen, hence can be considered equivalent methods of detection for split squeezed BECs.    

We observe that the range of entanglement that can be detected by the Giovannetti and covariance matrix methods is equivalently given by the total squeezing in the combined system. The entanglement in a single BEC can be witnessed from the Wineland squeezing parameter \cite{wineland1994}
\begin{equation}
\xi = \dfrac{N \Delta^2 [ S^{z^\prime}_{\text{tot}} ]}{|S^{x}_{\text{tot}}|^2 } , \label{Wineland}
\end{equation}
where
\begin{align}
S^{z^\prime}_{\text{tot}}  & = S^{z^\prime}_L + S^{z^\prime}_R \nonumber  \\
S^{x}_{\text{tot}} & = S^{x}_L + S^{x}_R
\end{align}
which quantifies the metrological usefulness of a state for Ramsey interferometry. For a single BEC, $\xi \ge 1$ for all separable states, $\xi<1$ reveals that the state is entangled. We find that the region that the Giovannetti and covariance matrix criteria detect entanglement coincides with all states with  $\xi<1$, as can be seen in Fig. \ref{criteriaDvsG}.  This is natural in view of the fact that the entanglement between the BECs originates from converting entanglement that is present originally in a single BEC to two BECs by splitting them.

\section{EPR steering} 
\label{sec:steering}

We show that our system allows for correlations stronger than entanglement, namely correlations allowing EPR steering \cite{EPR1935,ReidCOLLOQUIUM} 
The criterion we consider is the one proposed in \cite{ReidCOLLOQUIUM}, and experimentally adopted in \cite{fadel2017}, according which the left BEC steers the right BEC if there is a violation of the inequality
\begin{equation}
\mathcal{E}^{L\rightarrow R} \equiv \dfrac{ \sqrt{ \langle \Delta^2 [ g^{z} S_L^{z^\prime} - S_R^{z^\prime} )] \rangle \langle \Delta^2 [ g^{y} S_L^{y^\prime} - S_R^{y^\prime} )] \rangle  } }{ \langle S_R^x \rangle  } \geq 1 \;, \label{epr}
\end{equation}
where $g^{z}$ and $g^{y}$ are free real parameters that are chosen to minimize $ \mathcal{E}^{L\rightarrow R} $. 

\begin{figure}
\includegraphics[width=0.93\columnwidth]{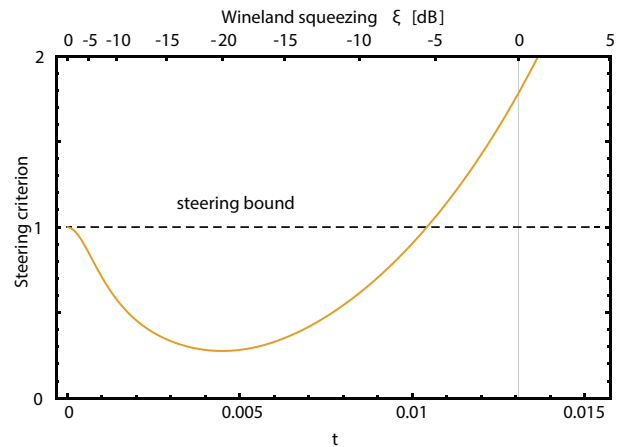}
\caption{EPR steering criterion (\ref{epr}) in a split squeezed Bose-Einstein condensate for $N=500$. The horizontal dashed line is the steerability bound, below which EPR steering is detected. The top scale shows the Wineland squeezing parameter (\ref{Wineland}). 
\label{criteriaEPR}
}
\end{figure}

In Fig. \ref{criteriaEPR} we plot the criterion (\ref{epr}). It is evident that the criterion can be violated for a wide range of $t$, and it is therefore suited to detect EPR steering between the two parts of a split spin-squeezed BEC.  The range is smaller than for entanglement since steering is a type of correlation that is stronger than entanglement, and for large times the type of correlation between the BECs cannot be described in terms of the steering variables that (\ref{epr}) is designed to be a witness of. The criterion (\ref{epr}) was successfully adopted in Ref. \cite{fadel2017} to detect steering between two regions of an expanded spin-squeezed BEC with $N\approx 600$.

We mention here that unlike entanglement or Bell correlations, steering is an intrinsically asymmetric concept: while one system can steer the other and vice-versa (two-way steering), there exists the possibility of having only one of the two systems able to steer the other (one-way steering) \cite{HePRL2015directional}. While in this theoretical study the state is symmetric, and therefore one-way steering always implies two-way steering, experimental implementation of our protocol might result in one-way steering only, because of asymmetric noise \cite{fadel2017}.

\section{Summary and conclusions} 
\label{v}

We have modeled and analyzed the behavior of split squeezed BECs. In our model, one spin polarized BEC is squeezed and split into two spatially separated parts. After the BECs are split, the particle number in each well collapses to a particular number state. The combination of the squeezing, splitting, and collapse is found to have the same effect as applying a Hamiltonian of the form $ H_{\text{eff}} = (S_L^z)^2 +  2 S_L^z S_R^z+(S_R^z)^2$, to two $ S^x $-polarized BECs with a total particle number of $ N $. This corresponds to a non-local entangling Hamiltonian combined with local squeezing on each well.  The situation found here is analogous to the generation of a two-mode squeezed state in optical systems.  Such a state is well-known to be equivalent to applying a beam splitter operation on single mode squeezed states \cite{schleich2011quantum}.  In the same way that photonic entanglement can be prepared by a beam splitter operation, here the entanglement between BECs is produced by splitting a single squeezed BEC into two spatially separated BECs.

The entanglement between the BECs was found to have the characteristic ``devil's crevasse'' form of entanglement which arises from a $ S_L^z S_R^z $ interaction, giving a non-zero entanglement for all time except the states equivalent to the initial product state. The marginal Wigner function, corresponding to the averaged quantum state over one BEC, evolves to a non-Gaussian state for evolution times exceeding $ t > 1/\sqrt{N} $. The conditional Wigner function, corresponding to a (post-selected) particle number collapse across the wells, shows a high degree of non-classicality, as illustrated by negativities appearing in the Wigner function. We also provided several approaches for the detection of entanglement using correlations of spin variables: the DGCZ, Giovannetti, and a generalized covariance matrix criteria. The Giovannetti and covariance matrix approaches were found to give the largest range of entanglement detection and appear to be best suited to detecting entanglement.

Using a small modification of the Giovannetti entanglement criterion, we showed that one may also detect EPR steering in the system.  An interesting question is whether Bell correlations between the two BECs could be detected by measuring low-order correlators, in a similar way. While there are known Bell inequalities that can be violated, these require always parity measurements which are experimentally challenging. Therefore, it is an intriguing theoretical task to find new Bell correlation witnesses that are more experimentally accessible for mesoscopic and macroscopic systems.

\begin{acknowledgments}
The authors thank Philipp Treutlein and Vinay Tripathi for discussions.  This work is supported by the Shanghai Research Challenge Fund; New York University Global Seed Grants for Collaborative Research; National Natural Science Foundation of China (61571301); the Thousand Talents Program for Distinguished Young Scholars (D1210036A); and the NSFC Research Fund for International Young Scientists (11650110425); NYU-ECNU Institute of Physics at NYU Shanghai; the Science and Technology Commission of Shanghai Municipality (17ZR1443600); and the China Science and Technology Exchange Center (NGA-16-001).
\end{acknowledgments}


\end{document}